\documentclass[sigconf, nonacm=true, authorversion=true]{acmart}

\usepackage{pifont}%
\newcommand{\cmark}{\ding{51}}%
\newcommand{\xmark}{\ding{55}}%
\AtBeginDocument{%
  \providecommand\BibTeX{{%
    \normalfont B\kern-0.5em{\scshape i\kern-0.25em b}\kern-0.8em\TeX}}}

\author[]{Henry Turner}
\email{firstname.lastname@cs.ox.ac.uk}
\affiliation{%
  \institution{University of Oxford}
  \city{Oxford}
  \country{UK}
}
\author{Simon Eberz}
\affiliation{%
  \institution{University of Oxford}
  \city{Oxford}
  \country{UK}
}
\author{Ivan Martinovic}
\affiliation{%
  \institution{University of Oxford}
  \city{Oxford}
  \country{UK}
}

\renewcommand{\shortauthors}{Turner et al.}

\usepackage{multirow}
\begin{document}

\title[Daily Turking]{Daily Turking: Designing Longitudinal Daily-task Studies on Mechanical Turk
}

\begin{abstract}
In this paper, we present our system design for conducting longitudinal daily-task studies with the same workers throughout on Amazon Mechanical Turk.
We implement this system to conduct a study into touch dynamics, and present our experiences, challenges and lessons learned from doing so.
Study participants installed our application on their Apple iOS phones and completed two tasks daily for 31 days.
Each task involves performing a series of scrolling or swiping gestures, from which behavioral information such as movement speed or pressure is extracted.

The completion of the daily tasks did not require extra interaction with the Mechanical Turk platform, yet paid workers through it.
This differs somewhat from the typical rapid completion of one-off tasks that workers are used to on Amazon Mechanical Turk.

This atypical use of the platform prompted us to evaluate aspects related to long-term worker retention and engagement over the study period, in particular the impacts of payment schedule (amount and structure over time) and reminder notifications.
We also investigate the specific concern of reconciling informed consent with workers' desire to complete tasks quickly.

We find that using the Mechanical Turk platform for conducting longitudinal daily task studies is a viable method to augment or replace traditional lab studies.

\end{abstract}

\maketitle
\section{Introduction}
Crowdsourcing has been embraced by academia and widely used in research tasks, being deployed to either collect data about workers (e.g., demographic surveys~\cite{paolacci2010running}), to collect data about some stimulus (e.g., image tagging~\cite{cocodataset}) or to collect data about workers reaction to stimuli (e.g., perceived audio quality~\cite{tacotron2}).
Commonly, these tasks are very short in duration and workers therefore attempt to make up for the small payment per task with volume~\cite{hara2018}.

In this work we explore the application of crowdsourcing to a new domain: %
collecting behavioral biometrics data via a longitudinal daily interactive study.
Behavioral biometrics -- the use of distinctive human behavior for identification or authentication -- have become a popular way to establish user identity without requiring explicit action (such as entering a PIN).
On mobile devices, touch dynamics involve using distinctive characteristics of touchscreen interactions such as the start and stop coordinates of swipes, gesture speed and touch pressure~\cite{frank2012touchalytics}.

Traditionally, studies on behavioral biometrics have been conducted in controlled lab environments~\cite{murmuria2015continuous,alghamdi2018dynamic,rocha2020continuous,frank2012touchalytics} due to the relative ease of setup and the physical presence of researchers to spot irregularities.
However, an ideal study requires a high number of participants to complete a specified phone-based task repeatedly and regularly over an extended period of time.
In addition, the makeup of participants should largely reflect the general population in terms of gender and, for touch dynamics, handedness.
Fulfilling these requirements is difficult for a lab study due to a limited local participant pool and the effort required to come to a lab daily.
In addition, restrictions related to the COVID-19 pandemic further impede in-person experiments, with safety requirements ruling out many potential experiments for the time being.

In order to avoid the shortcomings of lab studies, we undertook a mobile application based longitudinal study with recurring daily measurements on touch dynamics using the Amazon Mechanical Turk (MTurk) platform.
This allowed us to complete our data collection with high participation for a lower cost than in person experiments, whilst reducing COVID risks and the amount of researcher hours that have to be spent collecting data.

Using MTurk in this way is unconventional, and as far as we know studies of this kind have not been performed previously on the platform.
A key difference to typical MTurk studies is our requirement that tasks are performed on a daily basis \textit{by the same user} over an extended period.
Any data generated by workers who abandon the study or don't complete tasks for multiple days may be unusable for our subsequent work.
This requirement therefore makes both worker retention and continued worker engagement critical for our study success.

\vspace{0.3cm}
The main contributions of this work are:
\begin{itemize}
    \item We propose a system design for conducting interactive experiments with daily measurements that retain the same set of crowdsourced workers throughout.
    \item We evaluate the impacts of payment schemes, push notifications, worker engagement patterns, and worker attitudes on daily measurement studies.
    \item We explore concerns around informed consent specific to recruiting study participants on MTurk
    \item We provide design suggestions for those looking to complete similar studies using crowdsourced work platforms.
\end{itemize}

\section{Related Work}
In this section we discuss associated related work from previous research into crowdsourced work platforms, and conducting research studies on them, with a focus on MTurk.

\subsection{Typical MTurk Usage}
Several studies of both the MTurk platform and MTurk workers themselves have been conducted to gain insight into how the MTurk platform is used.

On the platform the majority of tasks completed (61\%) are microtasks~\cite{Hitlin2016}.
These tasks pay $\$0.10$ or less, can be completed quickly and are generally repetitive.
Examples of such tasks include image classification~\cite{Krishna2016} or relevance judgments~\cite{Alonso2011}.

In~\cite{Hitlin2016} tasks were observed for a week, with 37\% being imaged content tasks, 26\% audio transcription, 13\% content classification, 7\% information collection and 13\% completing surveys.
Our study does not fit into any of these categories, and is not a microtask, instead requiring longer continued engagement of specific workers, and thus is likely to be unfamiliar to workers.

\subsection{Longitudinal Studies on MTurk}
Previous longitudinal studies have been conducted on MTurk, with these generally being repeat surveys at pre-defined time intervals.
The TurkPrime platform supports longitudinal surveys, notifying workers of new surveys with emails~\cite{Litman2017}.
Using an email based notification strategy such as this yields participant re-response rates of 75\% after 2 months, decreasing to 47\% after 13 months, for a survey based task~\cite{Daly2015}.
Instead of email we elect to use mobile notifications on the device the worker is using to complete the experiment.
Previous works have used notifications to encourage continuous participation in experiments and in particular in experiments with high engagement frequency~\cite{smith2017developing, bidargaddi2018predicting}.
Notifications can also help establish habitual interaction with smartphones~\cite{Noe2019} and can be used to influence how often users access an app, in particular when offering a prize or electronic coupon~\cite{Hsu2020}.
In some sense our task is similar to this, as conducting additional measurements yields a direct financial reward.
In order to reinforce this feedback loop, we design our system such that payments for each measure are immediate, as described further in section~\ref{section:mturk_Design}.

Longitudinal studies have also been conducted to investigate information about workers over a period of time, such as that of Weinshel et al., who used a browser extension to monitor worker's exposure to website trackers over a period of a week, book-ending their study with surveys which prompted the install of a browser extension and a review of the data the extension had observed~\cite{Weinshel2019}. 
The key difference with our study is that interaction once the browser is installed with the study is passive until the completion of the post study survey.

In contrast to previous longitudinal studies, we sought to collect data from participants daily, and with a desire for reasonable certainty in participants continuing to complete the task each day.
Thus our task is both time critical, unlike most longitudinal studies, and requires the same workers participating, unlike most micro task studies.
As such, we design and implement our custom solution to conduct these high frequency longitudinal studies, as described later in Section \ref{section:mturk_Design}.

For micro-tasks that are repeated over long periods of time it has been demonstrated that task accuracy and success are both stable long term, with workers self-selecting out of tasks that they believe they perform poorly on~\cite{Hata2017}.
As such we can be confident that worker quality is unlikely to decline throughout the duration of our experiment, particularly as fatigue conditions are unlikely to accumulate significantly with the task only being completed once per day.

\subsection{Worker Recruitment \& Retention}
For many tasks on MTurk it is useful to retain workers so that a batch of tasks can be completed in a timely manner, with individual workers completing multiple tasks of the same type.
Typically the number of Human Intelligence Tasks (HITs) completed from a batch by the same worker follows a long tail distribution, with most workers completing few HITs, and a few workers completing most of the work, broadly aligning with the Pareto rule~\cite{Difallah2014, Hata2017}.
For our study retention is more than just useful: it is essential that the workers stay in the experiment for as long as possible.

Difallah et al. investigate the impact of different pricing schemes on different types of HIT, showing that pricing scheme impacts user retention for workers completing the same HIT repeatedly, with bonus payments for volume of HITs completed yielding the highest retention rates~\cite{Difallah2014}.
The schemes studied here can not be applied directly to our problem, but they help inspire the schemes we investigate.

Studies on task abandonment also show that insufficient monetary reward is the strongest driver of abandoning a task~\cite{Han2019}, with workers abandoning early if the monetary reward seems poor, and that payment is the strongest motivating factor for workers~\cite{Kaufmann2011, Rogstadius2011}.  
Workers also help each other using external platforms for mutual aid to flag good and bad requesters~\cite{McInnis2016, Han2020}, a further factor in ensuring that we pay workers appropriately, to ensure those using such services see our tasks as being worth completing.

The impact of rejection on workers pay has also been investigated~\cite{McInnis2016}, showing that workers are wary of rejected work, due to the loss of pay from wasted hours, as well as the reduced opportunities for future work due to worker filtering by rejection rate.
As such workers select tasks to minimize the risk of rejection and impact on their pay.
Similarly it has been demonstrated that longer instructions reduce task recruitment~\cite{wu2017confusing}.
This represents a difficulty for us, as the uniqueness (relative to the bulk of tasks) of our task, combined with our differing payment mechanisms for subsequent engagement may be perceived as a risk by workers and will necessitate longer instructions.

\section{Study Outline}
\begin{figure*}[t]
    \centering
    \includegraphics[width=\textwidth]{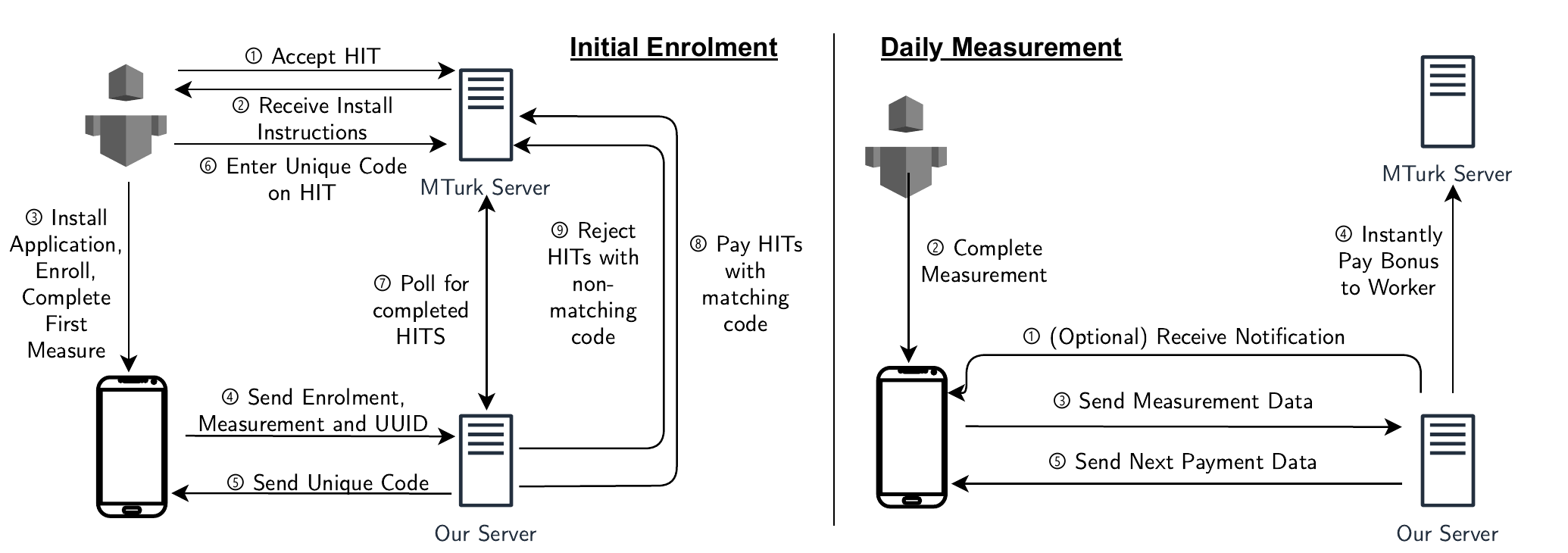}
    \caption{System diagram for our experiment platform. The left hand side of the figure shows the first enrollment task, whilst the right hand side shows the system operation for the remainder of the experiment period.}
    \label{fig:SystemDiagram}
\end{figure*}
We conducted our remote participant study with the purpose of creating a large datasets for behavioral touch based biometrics on smart phone devices. The study was approved by our institution's Institutional Review Board (reference XXX\footnote{blinded for review})

In using crowdsourced work to conduct our study, we had specific aims that we hoped to achieve throughout our experiment:
\begin{enumerate}
    \item[A1:] Collect measurements from workers daily, with few skipped days.
    \item[A2:] Engage workers for the full duration of the study, minimizing drop out rates.
\end{enumerate}
In this section we outline the daily task that workers were asked to complete, as well as our system design  through which we hope to achieve our two aims given above.

\subsection{Task Design}

Our biometrics measurement task aimed to capture touch based activity from users whilst they completed two separate tasks, scrolling through a news feed to find a specific article based on information given (similar in design to a Facebook news feed) and whilst counting the number of objects in a set of images, requiring repeat swipe gestures.

The two tasks were implemented in an iOS application, and would be presented to the user sequentially once they opened the application and began a measurement session.
We chose iOS as we need to standardize the device models used in our study for the eventual biometric data to be useful, and the number of iOS device models available is much smaller than the number of Android device models.
In a similar lab based study the same device would be used by all participants.

When outside of a measurement session the application showed them the reward for the next completion of the task, as well as the total value of rewards that could still be obtained for completions of the task in the study period.

Workers were asked to complete both the scrolling and swiping tasks once per day, in a process that takes approximately 4 minutes (median completion time 4 minutes 1.44 seconds), and for the full 31 day duration of the study period.

\subsubsection{Reminder Notifications}
\label{section:reminder_notifications}
Previous longitudinal studies with infrequent request intervals have made use of email to notify workers of new tasks to complete~\cite{Daly2015,Litman2017}.
However, these prior works had the benefit of being able to allow workers several days to complete tasks, whereas we sought to collect measurements daily for the study duration.
As a consequence we elected to use reminder notifications to improve user retention and encourage participants to complete the task each day, in the hopes of fulfilling both of our aims.

During the enrollment process we ask workers to enable notifications.
Upon submission of a successful measurement we schedule a reminder notification for the worker the subsequent day at 9 a.m. in their local timezone.
An additional notification is sent at 7 p.m. the same day if the worker has not submitted a measurement before this time.
We evaluate the effectiveness of these reminder notifications through our post task survey, as well as examining the times at which measurements were completed.

\subsubsection{Data Quality Assurance}
Quality data is a concern for any task on MTurk, but is particularly so in our case due to future studies being conducted with the data, and most other touch biometric works using supervised data collection methods.
For other tasks requesters use techniques such as obtaining multiple responses which can be combined later for classification problems, or including verification questions that the user must answer correctly to ensure they are human and attentive to the task~\cite{Mason2012}.
We employ a variant of the second of these methods, by validating the answers to each of the sub-tasks completed.

For the news feed scrolling sub-task if an incorrect article is selected then another round of the sub-task must be completed, until 5 rounds are answered correctly.
Similarly in the object counting sub-task if the incorrect number of each object is given then another round must be completed.
The correct article location is varied randomly for each round of the sub-task, and the number of objects, type of object, and number of images is varied for each round of the object-counting sub task.

\subsection{Mechanical Turk Design}

\label{section:mturk_Design}
In this subsection we describe the design of the MTurk related system components.

\subsubsection{Eligible Workers}
Our initial study recruitment HIT was available to all MTurk workers.
Whilst it is common to restrict studies to master workers, or those with a high HIT acceptance rates, we felt this was unnecessary for our task, as the barriers to entry, namely installing the app and completing the game for the measurement, would act as a filter on workers enrolling who were not motivated to complete the task.
Workers self-select to avoid risky tasks~\cite{McInnis2016}, and thus we were concerned that restricting participation to master workers may yield poor results, as our experiment differed enough from the platform norm that they may not see it as a worthwhile opportunity.

By ensuring the measurement task was designed robustly, and by requiring input from the touch screen, we believe that performing the tasks in the intended manner is the most efficient way to complete them.

\subsubsection{Infrastructure Design}
We designed our study infrastructure to complete all interactions with MTurk automatically, so that the study could run itself with minimal human oversight.
We conduct the completion of repeat tasks in a different way to previous longitudinal studies, in that only one HIT is completed by each worker, which we term the enrollment HIT.
After this HIT all additional work completed by the worker is paid using the MTurk bonus system.
This allows for faster payments (bonuses are paid instantly on task completion) and allows us to use out of band methods, specifically device notifications, for communicating with the workers and soliciting further work.
Furthermore the instant bonus payments help re-assure the workers that they will be paid for extra measurements they complete.
In the remainder of this section we detail the on-boarding process for workers who join our study, and the daily process completed by the workers in our study.

\subsubsection{On-boarding Process}
Recruitment was conducted by publishing a HIT to MTurk, which workers could claim and subsequently complete.
The initial task required workers to download the Testflight\footnote{https://developer.apple.com/testflight/} application onto an iOS device.
We limit sign-ups to select iOS devices to ensure similar devices are used for the biometric measurements, so that we can remove any device related effects when examining biometric traits at a later date.
This requirement was specified in the HIT title and description.
Whilst this could be added as a qualification to the HIT (Primary
Mobile Device - iPhone), it is unclear if all workers with access to an iPhone would have enabled this task requirement. In addition, the primary mobile device qualification does not allow requesters to specify a list of supported models as our study requires.

After installing our application the workers completed an enrollment process through the application.
This required them to read the informed consent documents, which were also included with the HIT.
Subsequently they then had to agree to a set of conditions based on providing consent, by toggling a switch for each condition.
The worker then provides us with basic demographic information about them, including country of origin, dominant hand, height and weight.

Lastly the worker completes the study task for the first time, receiving a verification code on completion.
The worker enters this code on the HIT page on MTurk.
Our backend infrastructure then validates this code, approving the HIT if it is correct, and associating the worker ID with the device that was given this code.
This verification step allows us to remove spam-like completions, or workers who attempted to guess the code and not complete the task.
A flow diagram of the initial on-boarding is shown in Figure~\ref{fig:SystemDiagram}.

\begin{table}[t]
\footnotesize
\begin{tabular}{crrrrrr}
\hline
\multirow{2}{*}{\textbf{\begin{tabular}[c]{@{}c@{}}Measures \\ Completed\end{tabular}}} & \multicolumn{3}{c}{\textbf{Pay Received (\$)}}                                                      & \multicolumn{3}{c}{\textbf{\begin{tabular}[c]{@{}c@{}}Equivalent\\  Hourly Pay (\$/h)\end{tabular}}} \\ \cline{2-7} 
                                                                                        & \multicolumn{1}{c}{\textbf{LC}} & \multicolumn{1}{c}{\textbf{HC}} & \multicolumn{1}{c}{\textbf{HI}} & \multicolumn{1}{c}{\textbf{LC}}  & \multicolumn{1}{c}{\textbf{HC}} & \multicolumn{1}{c}{\textbf{HI}} \\ \hline
1                                                                                       & 1.00                            & 1.00                            & 1.00                            & 7.50                             & 7.50                            & 7.50                            \\
11                                                                                      & 9.80                            & 12.30                           & 7.25                            & 12.19                            & 15.30                           & 9.02                            \\
21                                                                                      & 18.60                           & 23.60                           & 18.50                           & 12.61                            & 16.00                           & 12.55                           \\
31                                                                                      & 27.40                           & 34.90                           & 34.75                           & 12.77                            & 16.23                           & 16.20                           \\ \hline
\end{tabular}
\caption{Payment values for participants in the study. Equivalent hourly pay increases with participation duration. Time spent working was calculated by determining the median time taken to complete the daily task (241.44 seconds) and allowing 8 minutes for on-boarding.}
\label{tab:PaymentsData}
\end{table}
\subsubsection{Daily Measurements}
Each day workers are required to complete the same pair of tasks.
They can either do this unprompted, or can do it after being sent a reminder notification, the details of which were presented in Section \ref{section:reminder_notifications}.
On completing the two tasks the collected measurements are uploaded to our server.
The server responds by updating the in app values showing future potential earnings, and sends a pay bonus instruction to MTurk, to instantly pay the worker a bonus for the specified amount.
By instantly paying bonuses, and not requiring any manual approval, we hope to increase our workers confidence that they will be paid for this work completed outside of the normal HIT--payment structure.

The web server schedules the notifications for the next day at this point, assuming the participant has not finished the study.

\begin{figure*}[t]
    \centering
    \includegraphics[width=.9\textwidth]{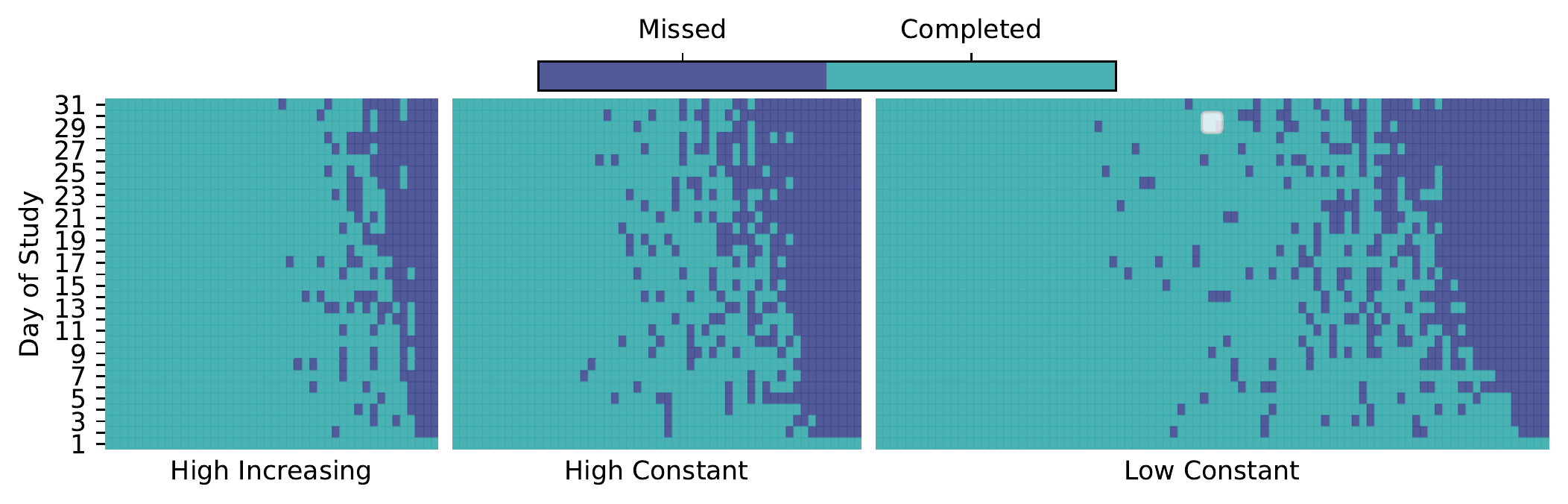}
    \caption{Heatmap showing measure completion by day for each participant in the study, split by payment scheme. Each vertical line represents a single worker's progress through the study.}
    \label{fig:CompletionHeatmap}
\end{figure*}

\subsubsection{Payment Schemes}
We utilized three different payment schemes within our study.
Two of the payment schemes, low constant (LC) and high constant (HC) paid the user a fixed amount for each measurement: $\$0.88$ and $\$1.13$ respectively.
The third scheme, high increasing (HI), paid users a variable amount, with each submitted measurement's value being determined by the formula:
\begin{equation}
    \text{Payment (cents)} = 40 + 5 \times (\text{\# Submitted Measurements - 1})
\end{equation}

Regardless of payment scheme, all participants were paid a fixed value of $\$1.00$ for the initial HIT for installing the application and completing the first measurement.
Table~\ref{tab:PaymentsData} shows the expected payment totals for each of the payment schemes with differing study completion percentages, as well as equivalent hourly pay calculated by examining the average completion times of workers on each of our tasks.
We note that the equivalent hourly pay for all of our tasks is a minimum of \$7.50 (when only the initial measure is completed), well above previously reported median worker hourly pay (\$3.18/h, calculated only on time completing tasks) and above the requester average of \$11.58/h~\cite{hara2018} for those who participate in the majority of the study.

We evaluate the impact of each of these payments schemes on overall participant engagement later in Section~\ref{section:payment_effect}.

\subsubsection{Soliciting Worker Feedback}
At the conclusion of our study we posted a survey HIT onto MTurk, to gather insights from the workers who performed work for us.
The survey asked questions specifically about the role of notifications, the value of payments, and how comfortable they were with completing studies like this using MTurk.

Unlike the other subtasks, we did not send a notification to workers to complete this as many workers had already uninstalled the app following study completion. We targeted our users through a qualification and offered a relatively high payment of \$1.00 to entice the workers to complete the survey.
We examine responses from this survey in several of the subsequent sections.

\section{Examining Study Outcomes}

In this section we analyze the feedback from our workers, and compare this with the activities of the workers within the experiment.
This allows us to gain an insight into the effectiveness of the mechanics we implemented to increase worker engagement, and to assess the suitability of MTurk for such experiments.

\subsection{Study Participants Statistics}
We enrolled a total of 187 participants (102 female, 82 male, 3 other) with the majority of users (153) from the US and the remaining 32 split over 16 other countries.
Of these, 44 were allocated to the HI payment scheme, 54 to HC and the remaining 89 to the LC scheme.
A further 3 workers submitted the initial measurement, but never associated the application with their MTurk account, and thus did not complete the enrollment.
While we initially recruited equal numbers of users into the HI and HC schemes, varying rates of rejected submissions (i.e., invalid verification codes) led to different numbers of users actually enrolled into the study.
These invalid codes could be either due to workers attempting to game the system (i.e., entering random codes in the hope of auto-approval) or workers accepting the HIT unaware of the restriction to iOS devices and submitting random codes instead of returning the HIT.

Overall most participants stayed engaged in the study, with the heatmap in Figure~\ref{fig:CompletionHeatmap} showing the proportion of participants who completed the study remained high until its conclusion, although with a gradual decline in participation throughout the period observed.
Overall 36.8\% of participants completed a measurement every day, and 68.4\% completed a measurement on more than 75\% of days.

The heatmap reveals certain trends in the participants who missed measurements, with vertical blue lines showing that often once a participant started missing measurements they were unlikely to re-engage with the study, including several users who exhibited this pattern after the initial sign up measurement.
This could be explained by participants who are trying to complete the task as fast as possible performing the on-boarding process, completing the measurement and subsequently removing the application, without realizing the potential future earnings.
This may be due to a lack of task clarity, as the unusual design of the task (vs. other typical MTurk HITs) means workers can't rely on previous knowledge, and they may have assumed subsequent HITs would be posted for subsequent measurements, or may not have understood that the app could be revisited for subsequent daily measurements, and removed it after submitting the HIT completion code.
Lack of clarity has been suggested in previous works as a cause of task abandonment~\cite{Han2019}

We also note some users show sporadic patterns, where they fail to complete the task on some days, but re-engage subsequently.
However, often missing a measurement proved terminal, with 46 of the 118 participants (39\%) who missed a measurement finishing the experiment with a run of three or more missed measurements.

\subsection{Varied Payment Schedules}
\begin{table*}[t]
\footnotesize
    \centering
\begin{tabular}{ccccccc}
 \toprule
       \textbf{Variable} & \textbf{Test} & \textbf{Scheme 1} &    \textbf{Scheme 2}   & \textbf{p-value} &\textbf{5\% sig.} &\textbf{10\% sig.} \\
 \midrule
 \multirow{3}{*}{\begin{tabular}{c}
       Drop Out  \\
       After First
  \end{tabular}} &  \multirow{3}{*}{\begin{tabular}{c}
       Two Proportion  \\
       Z-Test
  \end{tabular}} 
     & HI & HC & 0.464 & \xmark & \xmark \\
   & & HI & LC & 0.572 & \xmark & \xmark \\
   & & HC & LC & 0.133 & \xmark & \xmark \\ \midrule
 
\multirow{3}{*}{\begin{tabular}{c}
       Measures  \\
       Completed
  \end{tabular}} &  \multirow{3}{*}{\begin{tabular}{c}
       Two-Tailed  \\
       t-Test
  \end{tabular}}
     & HI & HC & 0.131 &\xmark &\xmark \\
   & & HI & LC  & 0.122 & \xmark & \xmark \\
   & & HC & LC  & 0.870 & \xmark & \xmark \\ \midrule
 
  \multirow{3}{*}{\begin{tabular}{c}
       Measures  \\
       Completed
  \end{tabular}} &  \multirow{3}{*}{
  \begin{tabular}{c}
       One-Tailed t-Test  \\
       $\text{Scheme }1>\text{Scheme }2$
  \end{tabular}} 
     & HI & HC & 0.065 & \xmark & \cmark \\
   & & HI & LC & 0.061 & \xmark & \cmark \\
   & & HC & LC & 0.565 & \xmark & \xmark \\ \midrule
   
   \multirow{3}{*}{\begin{tabular}{c}
       Completed  \\
       Every Day
  \end{tabular}} &  \multirow{3}{*}{\begin{tabular}{c}
       Two Proportion Z-Test \\
       Scheme 1 > Scheme 2
  \end{tabular}} 
     & HI & HC & 0.019 & \cmark & \cmark \\
   & & HI & LC & 0.014 & \cmark & \cmark \\
   & & HC & LC & 0.554 & \xmark & \xmark \\ 
   
   \bottomrule \end{tabular}
    \caption{Statistical significance test results for varying payment schedules. Measures completed tests are on the number of measures completed per user. Drop out after first are based on proportion of users who only complete one measurement and never return. Number of samples are HI:44, HC:54 and LC:89.}
    \label{tab:payment_stats}
\end{table*}
\begin{figure*}[t]
\centering
\begin{minipage}{.49\textwidth}
 \centering
    \includegraphics[width=.8\textwidth]{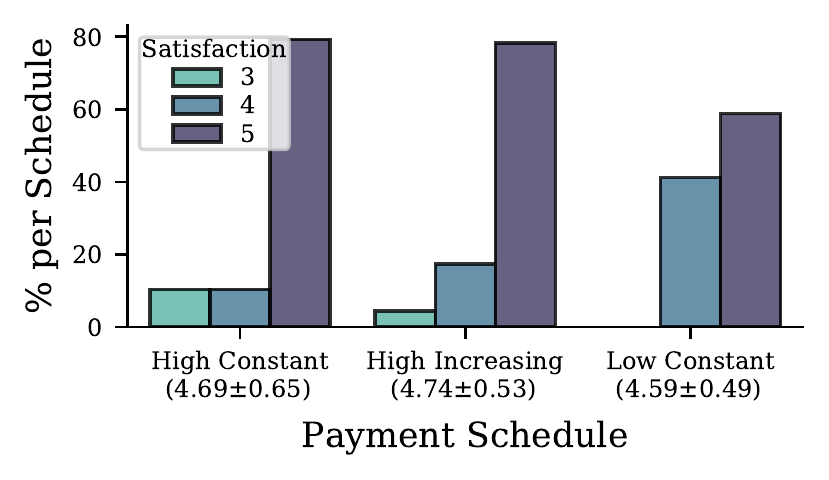}
    \caption{Respondent's payment satisfaction, split by payment schedule}
    \label{fig:SchedulevsSat}
\end{minipage}
\hfill
\begin{minipage}{.49\textwidth}
 \centering
    \includegraphics[width=.78\textwidth]{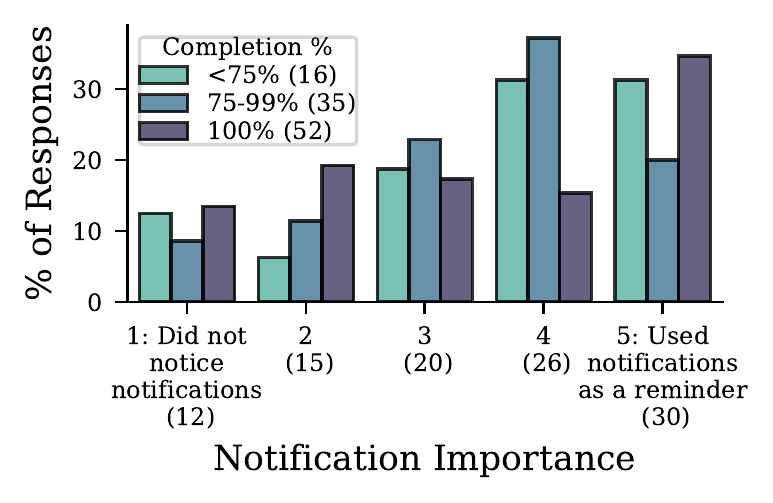}
    \caption{Self-reported importance of notifications, broken down by completion rate}
    \label{fig:NotificationImportance}
\end{minipage}
\end{figure*}
We evaluate the payment schedules used in our experiment to determine any impacts they have on completion rate, as well as payment satisfaction compared with completion rate and with payment schedule.

We first investigate the patterns in completion between the three payment schedules, investigating if the differences in: i) participants who only completed 1 measure, ii) mean measures completed (excluding those who dropped out after one) and iii) completing all measurements, are statistically significant between the pairs of payment schemes.

P-values for these comparisons are reported in Table~\ref{tab:payment_stats}.
It is clear that there is no significant different between payment schemes on the number of users who complete only one measurement.
This suggests that not continuing with the experiment may not be related to payment, and is more likely to be failing to understand the daily nature of repeat tasks through the application.

Examining the total measures completed per user, we find that with a two tailed test there is no significant difference between the payment schemes.
We conducted a one tailed test to examine if the increasing scheme leads to more measures, and if the higher constant has an impact, but we find no significant difference at the 5\% confidence level.
Further experiments are necessary to further investigate this effect, particularly with a larger sample size to enable better confidence in the values.

Examining the proportion of participants who completed the task daily, we find that there is a statistically significant difference showing the HI scheme has a higher proportion than the two constant schemes.
This supports our theory that by participants observing the increasing payments over time, and seeing that their future earnings are dependent on completing it daily, they are more motivated to complete the task fully and not miss sessions, as this would impact future earnings.

\label{section:payment_effect}
We also investigate the impact on workers payment satisfaction with respect to their payment schedule through our post-study survey.
Figure~\ref{fig:SchedulevsSat} shows the payment satisfaction depending on the payment schedule.
We observed mean scores of 4.74, 4.69 and 4.59 for the HI, HC and LC schedules, with none of the differences being statistically significant.

An overall explanation for the lack of significant differences here could be that all of our payment schemes are notably higher than the worker average (and above US minimum wage).
Consequently most ratings given are high, and that workers are satisfied with complete the study on any of our payment schemes.
Further experiments could be conducted to determine lower wage levels where an impact is seen, however it is not ethical to pay workers poorly, so care must be taken in the design of these.
The value of determining a minimum bound for pay is also ethically questionable, as it will facilitate the further exploitation of workers with low wages.

\subsection{Daily Completion Reminders}
\begin{figure*}[t]
\centering
    \includegraphics[width=1\textwidth]{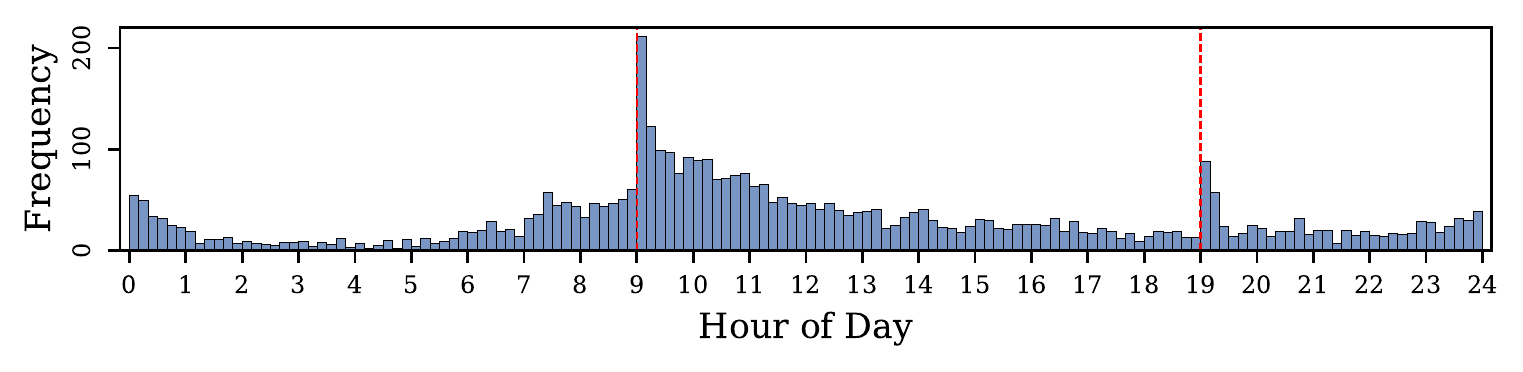}
    \caption{Submission times in user's timezone. Spikes can be observed directly following reminder notifications at 9~a.m. and 7~p.m (shown by red dashed lines). Bins spaced at ten minutes interval.}
    \label{fig:completiontime}

\end{figure*}
In our post study survey participants were asked ``How important were the daily notifications in reminding you to complete the task?'', and required to provide an answer on a scale from 1 (Completed the task without noticing the notifications) to 5 (Used the notifications as a reminder to complete the daily task).
Figure~\ref{fig:NotificationImportance} shows that 54\% of respondents rated the importance of notifications high (4 or 5) but that the self-reported importance of notifications had no statistically significant effect on actual completion rates.
We also looked to see if there was any difference between those who did not enable notifications when registering and those who did, however due to a low sample size not having notifications (22 workers) we could not draw any conclusions.

To further evaluate the actual rather than perceived importance, we also recorded the completion time of each submitted sample (see Figure~\ref{fig:completiontime}).
The data shows spikes in the ten minute windows following notification dispatch (9 a.m. and 7 p.m.) followed by a trailing off effect after these notifications.
While it could be argued that both are popular times to complete these tasks in general, the size of the spike in specific 10-minute windows strongly  suggests a significant impact of the notification timing on prompting workers to complete the task at these times.

\section{Remote studies and informed consent}
Academic research involving human subjects is subject to IRB approval at most institutions and researchers are typically required to follow a formal ethics procedure involving participants giving informed consent.
As part of this process, we provided a project information sheet in the HIT and within the app along with the app-based consent form.
As part of the consent form, participants indicate to have read and understood the information sheet and that they have received answers to any questions they may have had.

\begin{figure}
    \centering
    \includegraphics[width=\columnwidth]{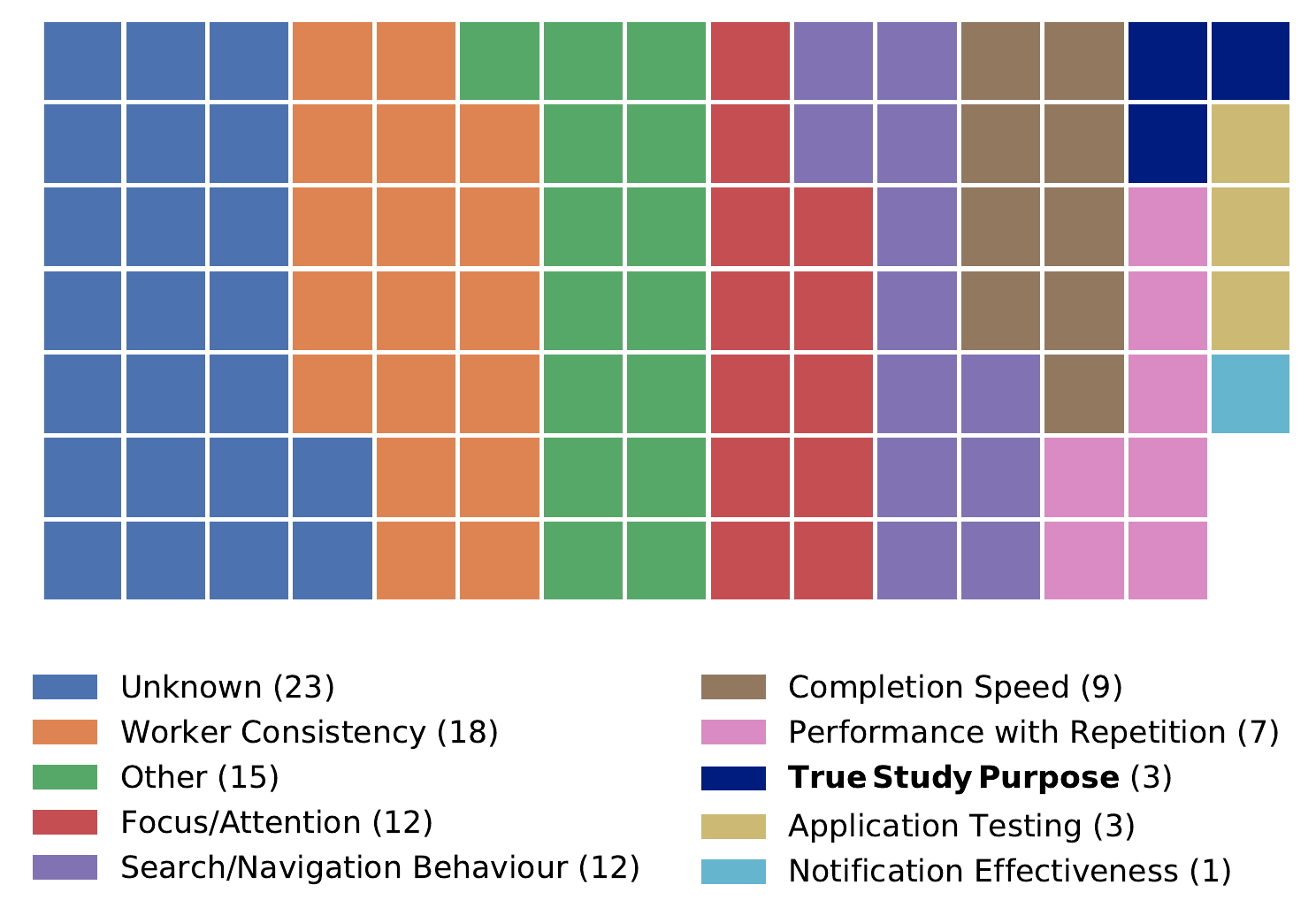}
    \caption{Categorized responses given by participants when asked about the study purpose. }
    \label{fig:WafflePlot}
\end{figure}

During the study's exit survey, we asked participants what they thought the purpose of the study was\footnote{By `the study' we mean the touch authentication study that workers completed, as opposed to this subsequent study of worker behaviour.}.
The study purpose was clearly stated in the first paragraph of the information sheet.
Figure~\ref{fig:WafflePlot} shows the responses given by participants.
Only 3 respondents' (2.91\%) answers related to the true study purpose.
The remaining users either stated that they did not know or gave various guesses based on the tasks (such as attention span, consistency or improvement over time).

While the time gap between enrollment in the study (at which point the information sheet is presented) and exit survey is over a month, the survey results nevertheless suggest that only a small fraction of participants read and understood the project information sheet.
This result highlights a severe disconnect between the realities of MTurk and academic due process.
Due to the relatively small payment for the average HIT, workers are primed to complete them quickly with minimal delay in order to achieve a good hourly rate of payment.
Naturally, this emphasis on speed is opposed to giving informed consent as workers are unlikely to consider reading information sheets worth their time.

While it could be argued that the present study is relatively low-risk as it does not involve collecting sensitive data, this fact would arguably be unknown to participants without reading the information sheet as HITs do not undergo any review process before going live.
An additional factor is likely the high fraction of HITs posted by companies not required to follow this process, leading to users not being conditioned to seeing consent forms in general.

Our results show that the traditional academic process of informed consent is not fit for purpose on MTurk and does not lead to truly informed consent for the vast majority of participants.
For low-risk experiments (such as ours) it could be argued that it is sufficient to give participants  the \textit{opportunity} to thoroughly inform themselves without strong verification that they really did but this is less true for riskier studies (e.g., those involving sensitive information).
While typical high-risk experiments (such as medical trials) are often limited to the physical domain and rare on MTurk, it is still important to consider options for obtaining truly informed consent.
Standard approaches to verify that users have read the information sheets (such as quiz or trapping questions) add complexity and significantly increase time spent on the process.
This is particularly true for common tasks of short duration as the time spent on consent would heavily outweigh the actual participation time. 
One caveat to this is that we did not require our workers to be ``Master workers'' or to meet a work completion threshold, and as such it may be that introducing requirements on worker quality would change this outcome.

\section{Future Design Suggestions}
In light of our experimental results and our experiences running this study, we use this section to distill some design suggestions for those looking to build similar systems in the future.

\subsubsection*{Notifications for Retention} Our results suggest that the notifications helped many of our workers stay engaged in the experiment.
We observed large spikes in tasks being completed immediately following a notification, and user feedback suggested they were useful for many workers.
Thus we would suggest using a notification scheme for any future app based studies.
For a non-app based study we would suggest either using email notifications, or browser notifications.

Finally we would also suggest improving over our system by sending notifications to re-engage users who have been active but then miss a measurement.
In our system this led to no more notifications being sent, but our results suggests that otherwise active workers may miss a measurement and then no longer engage.
Sending another notification the day after missed measurements may help re-engage these lost workers.

\subsubsection*{Design Payment Schemes Carefully} Our payment scheme experiments showed that our increasing payment scheme led to a significantly higher proportion of users completing the entire study, whilst we found no significant difference between the payment scheme and mean number of submitted measures.
As such we would recommend using an increasing scheme, but can not recommend an overall payment level.
In our study we paid workers significantly more than the average, as we felt that paying them wages below minimum wage is ethically wrong.
It may be that workers would perform the task just as well on lower overall wages, and that the improvements from increasing the payment with each measure would persist at lower overall payment levels.

\subsubsection*{Take care with informed consent} On surveying our workers post-study, worryingly few of them still remembered the aim of the study.
Whilst it may be that this would improve by filtering worker quality on sign up more aggressively, it may be that the combination of longer duration task than normal and workers moving quickly, that few of them were truly informed when they consented.
Thus we would suggest that researchers should verify whether participants' consent was truly informed and that the degree of this verification should be proportionate to the risk level of the experiment.

\subsubsection*{Automate as much as possible} In our system we automate the initial HIT approval, the notification sending, and all bonus payments each time a measurement is submitted. 
High levels of automating make it much easier to scale the number of participants in the study, and allow the study to take place without daily involvement of those running the study.
Anecdotally several workers told us in our post study survey that they appreciated the speed with which bonus payments were sent on task completion.

\subsubsection*{Design your task with work quality in mind} As in other work previously conducted on MTurk, it is important to ensure the quality of the data submitted through the platform.
In our case the games required touch input on a mobile device to be completed, making any automatic completion system harder to use, and further to this required workers to get the answers to each task round correct in order to proceed.
When designing a similar high repeat system, we would suggest employing similar designs, in particular because it is not possible to rescind the bonus payments once they are sent, so workers completing fraudulent work will be able to get paid for completing the task if they are improperly designed.

\section{Conclusion}
Overall, MTurk is a viable platform for long-term studies that require daily completion, and the participant retention rate is high, with 37\% of users completing all 31 samples and 68\% completing at least 75\%.
We would recommend the platform for conducting studies similar to ours, and think that by designing the task well, and designing for worker retention and engagement, high quality datasets can be obtained.

Examining payment schedules, we observed no statistically significant relationship between payment schedule and payment satisfaction or satisfaction and average completion rate.
Likewise there was no statistically significant difference between payment schemes on the number of measurements completed, or the number who only completed the initial measurement. %

However, we did find that our payment schedule which increases the payment with each measurement had a statistically significant effect on the proportion of workers who completed all measurements.
As such we would recommend future studies implement similar mechanisms in their payment schedules if workers completing the full study is of importance.

Twice-daily push notifications were a useful tool to improve retention, with our data showing large spikes in submissions shortly after reminders notifications are sent.

The demographics obtained in our study without explicit targeting is largely comparable to previous work, showing a roughly equal gender split and a largely US-based set of participants. 
Our exit survey revealed that only 2.9\% of participants read, understood and retained the contents of the participant information sheet, suggesting that the standard academic approach for obtaining informed consent may not be compatible with workers' focus on completing HITs as fast as possible.
The general feedback received from workers was positive and our results show that MTurk is a highly viable tool to augment or replace lab studies in fields going beyond the traditional scope of the platform.

\section{Acknowledgments}
This work was generously supported by a grant from Mastercard and by the Engineering and Physical Sciences Research Council (grant number EP/P00881X/1). We would also like to thank Martin Georgiev and Giulio Lovisotto who were both involved in the original study into touch dynamics.

\bibliographystyle{ACM-Reference-Format}
\bibliography{refs}

\end{document}